\documentclass[preprint,aps,floatfix,nofootinbib,superscriptaddress]{revtex4}

\usepackage{graphicx}
\usepackage{amsmath}
\usepackage{amssymb}
\usepackage{longtable}

\newcommand{\rd}{\partial}
\newcommand{\1}{\mbox{1}\hspace{-0.25em}\mbox{l}}

\newcommand{\scalar}{\bar{\psi}\psi}
\newcommand{\dt}{\Delta t}
\newcommand{\dts}{{\Delta t}_s}
\newcommand{\NssN}{\left\langle N|\bar ss|N\right\rangle}
\newcommand{\NllN}{\left\langle N| \bar{u}u + \bar{d}d | N\right\rangle}

\begin{document}


\begin{flushright}
\normalsize
UTHEP-615        \\
KEK-CP-241       \\
\end{flushright}

\title{
Nucleon strange quark content 
from two-flavor lattice QCD with exact chiral symmetry
}

\newcommand{\Tsukuba}{
  Graduate School of Pure and Applied Sciences, University of Tsukuba,
  Tsukuba, Ibaraki 305-8571, Japan
}
\newcommand{\CCS}{
  Center for Computational Sciences, University of Tsukuba, Tsukuba, 
  Ibaraki 305-8577, Japan
}
\newcommand{\KEK}{
  KEK Theory Center,
  High Energy Accelerator Research Organization (KEK),
  Tsukuba 305-0801, Japan
}
\newcommand{\GUAS}{
  School of High Energy Accelerator Science,
  The Graduate University for Advanced Studies (Sokendai),
  Tsukuba 305-0801, Japan
}
\newcommand{\Osaka}{
  Department of Physics, Osaka University,
  Toyonaka 560-0043, Japan
}

\author{K.~Takeda}
\affiliation{\Tsukuba}

\author{S.~Aoki}
\affiliation{\Tsukuba}
\affiliation{\CCS}

\author{S.~Hashimoto}
\affiliation{\KEK}
\affiliation{\GUAS}

\author{T.~Kaneko}
\affiliation{\KEK}
\affiliation{\GUAS}

\author{J.~Noaki}
\affiliation{\KEK}

\author{T.~Onogi}
\affiliation{\Osaka}

\collaboration{JLQCD collaboration}
\noaffiliation

\date{\today}

\begin{abstract}
  The strange quark content of the nucleon 
  $\langle N | \bar{s}s | N \rangle$
  is calculated in dynamical lattice QCD employing the overlap fermion
  formulation.
  For this quantity, exact chiral symmetry guaranteed by the
  Ginsparg-Wilson relation is crucial to avoid large contamination due
  to a possible operator mixing with $\bar{u}u+\bar{d}d$.
  Gauge configurations are generated with two dynamical flavors on a
  $16^3 \times 32$ lattice at a lattice spacing $a\simeq 0.12$~fm.
  We directly calculate the relevant three-point function on the
  lattice including a disconnected strange quark loop
  utilizing the techniques of the all-to-all quark propagator and low-mode
  averaging. Our result
  $f_{T_s}\!=\! m_s \NssN / M_N \!=\! 0.032(8)_{\rm stat}(22)_{\rm sys}$,
  where $m_s$ and $M_N$ are strange quark and nucleon masses,
  is in good agreement with our previous indirect estimate using the
  Feynman-Hellmann theorem.
\end{abstract}

\pacs{}

\maketitle
\section{Introduction}

In the naive quark model, the nucleon consists of three valence up and
down quarks. 
This picture is made more precise by taking account of quantum
effects based on quantum chromodynamics (QCD), the fundamental theory
of strong interaction, with which one expects additional effects due
to the gluon and sea quark degrees of freedom. 
In fact, in high energy hadron scatterings, these effects are observed
as parton distributions of the gluon and sea quarks, which can be
analyzed using perturbative calculations of QCD. 
At low energy, quantitative calculation of the sea quark effect is far
more difficult because of the nonperturbative nature of QCD. 
In this work, we consider the nucleon strange quark content 
$\langle N|\bar{s}s|N\rangle$.
This matrix element directly measures the effect of sea quark, because
there is no valence strange quark in the nucleon. 

The nucleon strange quark content represents the effect of strange 
quark on the mass of the nucleon, which is often parametrized by 
\begin{equation}
  f_{T_s} = \frac{m_s \NssN}{M_N},
  \label{eqn:fts}
\end{equation}
where $m_s$ and $M_N$ are the masses of strange quark and nucleon. 
$\bar{s}s$ is the scalar operator made of strange quark fields. 
This parameter is also relevant to the dark matter searches, as one of
the candidates ---neutralino in the supersymmetric models--- may
interact with the nucleon most strongly through its strange quark
content via the Higgs boson exchange diagram
\cite{Griest:1988ma,Drees:1993bu,Bottino:2001dj,Baltz:2006fm,Ellis:2008hf}.  
The magnitude of the matrix element $\langle N|\bar{s}s|N\rangle$ is
therefore directly related to the sensitivity of the present
\cite{Angle:2007uj,Ahmed:2009zw} and future experiments. 

Another quantity of physical interest is a ratio of strange quark and
light (up and down) quark contents: 
\begin{equation}
  y \equiv \frac{2\NssN}{\NllN}.
   \label{eqn:y}
\end{equation}
The denominator $\langle N|\bar{u}u+\bar{d}d|N\rangle$ corresponds to
the nucleon $\sigma$ term, which is relatively well-determined as it
is related to an amplitude of the pion-nucleon scattering. 
This is not the case for $\langle N|\bar{s}s|N\rangle$, for which only
lattice QCD can potentially make a quantitative prediction. 

\begin{figure}[tbp]
 \centering
 \includegraphics[width=0.4\textwidth,clip]{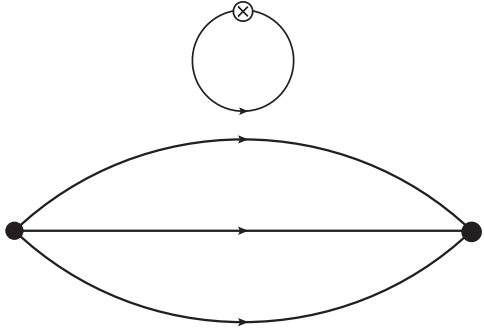}
 \caption{
   Disconnected three-point function relevant to $\NssN$.
   Lines show quark propagators that are dressed by virtual gluons
   and sea quarks in QCD.
   The connected three lines correspond to the nucleon propagation
   and the disconnected loop arises from the strange scalar operator
   $\bar{s}s$. 
 }
 \label{fig:diagram}
\end{figure}

The numerical calculation of the matrix element 
$\langle N|\bar{s}s|N\rangle$ on the lattice is not straightforward,
because it involves a disconnected quark-loop diagram shown in
Fig.~\ref{fig:diagram}. 
With the conventional method to calculate the quark propagator in
lattice QCD, the computational cost to obtain the disconnected 
quark loop is prohibitively high, as one has to perform an
expensive inversion of the Dirac operator for sources located at every lattice
sites; 
the computational cost is then proportional to the lattice volume
squared, $(N_s^3)^2$. 
Furthermore, since the scalar operator $\bar{s}s$ may have nonzero
vacuum expectation value (VEV), which is divergent when $m_s$ is
finite, one has to subtract this VEV contribution to extract the
physical matrix element $\langle N|\bar{s}s|N\rangle$.
This requires a large cancellation that induces a large statistical
error. 

In this work, we overcome these practical difficulties in the lattice
calculation by using the methods of the low-mode averaging
\cite{DeGrand:2004qw,Giusti:2004yp} and the all-to-all propagator
\cite{Foley:2005ac,Bali:2005fu}.
The all-to-all propagator allows us to calculate the propagation of
the quark between arbitrary lattice sites at once, by introducing a 
stochastic estimator (for a practical implementation, see below). 
Although it introduces additional statistical noise, the low-mode
averaging eliminates the noise for physically relevant low-lying
quark-mode contributions and improves the statistics by averaging over 
space-time lattice sites. 
These techniques are crucial for the calculation of the disconnected
diagram in lattice QCD.

Another important advantage of this work over the previous lattice
calculations of $\langle N|\bar{s}s|N\rangle$ 
\cite{Fukugita:1994ba,Dong:1995ec,Gusken:1998wy,Michael:2001bv}
is the use of a lattice fermion formulation that preserves exact chiral
symmetry at finite lattice spacings. 
For both sea and valence quarks we employ the overlap fermion
\cite{Neuberger:1997fp,Neuberger:1998wv}, 
which satisfies the Ginsparg-Wilson relation \cite{Ginsparg:1981bj}
and thus has a symmetry under a modified chiral transformation
\cite{Luscher:1998pqa}. 
This exact chiral symmetry prohibits the operator mixing  under the
renormalization between $\bar{s}s$ and $\bar{u}u+\bar{d}d$, where the
matrix element of the latter operator involves the connected diagram
contribution. 
With the Wilson fermion formulation that has been used in the
previous works, the operator mixing is induced due to the explicit
chiral symmetry breaking on the lattice.
Since the connected diagram contribution of $\bar{u}u+\bar{d}d$ is
larger than the disconnected one by an order of magnitude, this may
give rise to a large systematic error unless the mixing contribution is
subtracted nonperturbatively.

In our previous work \cite{Ohki:2008ff}, we used a technique to extract
$\langle N|\bar{u}u+\bar{d}d|N\rangle$ and 
$\langle N|\bar{s}s|N\rangle$ 
from the quark mass dependence of the nucleon mass using the
Feynman-Hellman theorem.
Since the number of sea quark mass values in the simulations was
limited, the method had an inconsistency that 
the disconnected contribution was evaluated at up and down quark
masses, which are different from the physical strange quark mass.
In the present work, this limitation no longer remains.
Although the calculation is done on two-flavor QCD lattices, which are
available from the project of the dynamical overlap fermion by the
JLQCD-TWQCD Collaboration \cite{Aoki:2008tq}, an extension to the
realistic 2+1-flavor QCD is straightforward and in fact underway.

This paper is organized as follows.
In Sec.~\ref{Sec:Simulation}, our simulation setup and the methods
of the all-to-all propagator and the low-mode averaging are described.  
We investigate the efficiency of the low-mode averaging by comparing
the statistical error of the nucleon two-point function as presented 
in Sec.~\ref{Sec:effective}.
Extraction of the strange quark content from the disconnected
three-point function is discussed in Sec.~\ref{Sec:Results}.
Section~\ref{Sec:Extrap.} is devoted to a discussion of chiral
extrapolation to the physical quark masses. 
In Sec.~\ref{Sec:Renorm.}, we emphasize an important role 
of chiral symmetry in the calculation of the strange quark content.
We also make a comparison with previous works including the recent
results \cite{Young:2009zb,Toussaint:2009pz}.
Our conclusions are given in Sec.~\ref{Sec:Conclusion}.
A preliminary report of this work is found in \cite{Takeda:2009ga}.

\section{Simulation details}
\label{Sec:Simulation}

\subsection{Simulation Setup}
On a four-dimensional Euclidean lattice we simulate QCD with two
flavors of degenerate up and down quarks.
As the lattice formulation, we use the Iwasaki gauge action and the
overlap quark action. 
The overlap-Dirac operator is given by
\cite{Neuberger:1997fp,Neuberger:1998wv} 
\begin{equation}
  D(m)
  = 
  \biggl(m_0+\frac{m}{2}\biggr)+\biggl(m_0-\frac{m}{2}\biggr)
  \,\gamma_5\,{\rm sgn}\left[ H_W \right],
  \label{eqn:ov}
\end{equation}
where $H_W=\gamma_5 D_W(-m_0)$ is the Hermitian Wilson-Dirac operator and 
$m_0=1.6$ in this study.
The mass parameter $m$ corresponds to the up-down or strange quark mass.
We also introduce an additional Boltzmann factor \cite{Fukaya:2006vs}
which does not change the continuum limit of the theory 
but substantially reduces the computational cost to calculate 
${\rm sgn}[H_W]$ by prohibiting the exact zero modes and suppressing
near-zero modes of $H_W$.  
This additional Boltzmann factor induces a side effect that the 
{\it global} topological charge $Q$ during the hybrid Monte Carlo update
is fixed.   
We simulate only the trivial topological sector $Q=0$ in this study;
the effect of fixing topology is suppressed 
by an inverse power of the space-time volume $1/(N_s^3 N_t)$
\cite{Aoki:2007ka} 
and turns out to be small (typically below a few percent level)
in our studies of meson observables 
\cite{Aoki:2008ss,Noaki:2008iy,Aoki:2009qn}.
We expect that it is even smaller for baryons.

Our gauge configurations are generated 
on a $N_s^3\times N_t=16^3\times 32$ lattice 
at a gauge coupling $\beta=2.30$ where 
the lattice spacing is determined as $a=0.118(2)$~fm 
using the Sommer scale $r_0 =0.49$~fm as an input.
We accumulate 100 independent configurations of two-flavor QCD
at three values of up and down quark masses 
$m_{ud}=0.025$, 0.035, and 0.050, 
which cover a range of the pion mass $M_{\pi}=$ 370--520 MeV.
The physical quark masses are fixed as 
$m_{ud,\rm phys}=0.0034$ and $m_{s,\rm phys}=0.077$ 
from our analysis of the pion and kaon masses \cite{Noaki:2008iy,m_s}.
We refer the readers to \cite{Aoki:2008tq} 
for further details of the configuration generation.

We take two values of the valence strange quark mass 
$m_{s,val}=0.070$ and $0.100$ close to $m_{s,\rm phys}$, 
and calculate two- and three-point functions 
\begin{eqnarray}
   C_{2\rm pt}^\Gamma({\bf y},t_{\rm src},\dt) 
   & = & 
   \frac{1}{N_s^3} 
   \sum_{\bf x} \mathrm{tr}_s
   \left[ 
      \Gamma \langle 
                N({\bf x},t_{\rm src}+\dt)\bar{N}({\bf y},t_{\rm src})
             \rangle
   \right], 
   \label{2pt}
\\
   C_{3\rm pt}^\Gamma({\bf y},t_{\rm src},\dt,\dts)
   & = &
   \frac{1}{N_s^6} 
   \sum_{\bf x,z}
   \left\{
      \mathrm{tr}_s \left[
         \Gamma \langle 
                   N({\bf x},t_{\rm src}+\dt)S^{\rm lat}({\bf z}, t_{\rm
		   src}+\dts)
                   \bar{N}({\bf y},t_{\rm src})
                \rangle
      \right]
   \right.
\notag \\
   && 
   \left.
      \hspace{7mm}
     - \langle S^{\rm lat}({\bf z},t_{\rm src}+\dts) \rangle \,
      \mathrm{tr}_s\left[ 
         \Gamma \langle 
                   N({\bf x},t_{\rm src}+\dt) \bar{N}({\bf
		   y},t_{\rm src})
         \rangle
      \right]
   \right\},  
\label{3pt}
\end{eqnarray}
where we use the nucleon interpolating field 
$N=\epsilon^{abc}(u_a^T C\gamma_5d_b) u_c$ with the charge conjugation
matrix $C=\gamma_4\gamma_2$.
The trace ``$\mathrm{tr}_s$'' is over spinor index of the
valence nucleon
and $\langle\cdots\rangle$ represents a Monte Carlo average.
The scalar operator made of the strange quark field is given by
\begin{equation}
  S^{\rm lat}=\bar{s}\biggl(1-\frac{D(0)}{2m_0}\biggr)s  
\end{equation}
on the lattice for the overlap-Dirac operator (\ref{eqn:ov}).
To obtain the continuum operator $S^{\rm cont}(\mu)$ at the energy scale
$\mu$, we need the renormalization factor $Z_S(\mu)$ as 
$S^{\rm cont}(\mu)=Z_S(\mu)S^{\rm lat}$.
The details including possible operator mixing are discussed in
Sec.~\ref{Sec:Renorm.}.

We take two choices of the projection operator
$\Gamma\! =\!\Gamma_{\pm}\!=\! (1\pm\gamma_4)/2$,
which correspond to 
the forward and backward propagating
nucleons, respectively. 
The two- and three-point functions are averaged over 
the two choices of $\Gamma$ 
\begin{eqnarray}
   C_{2\rm pt}({\bf y},t_{\rm src},\dt)
   & = & 
   \frac{1}{2}
   \left\{
      C_{2 \rm pt}^{\Gamma_{+}}({\bf y},t_{\rm src},\dt)
     +C_{2 \rm pt}^{\Gamma_{-}}({\bf y},t_{\rm src},N_t-\dt)  
   \right\}
   \label{2pt.average}
   \\
   C_{3 \rm pt}({\bf y},t_{\rm src},\dt,\dts)
   & = &
   \frac{1}{2}
   \left\{
      C_{3 \rm pt}^{\Gamma_{+}}({\bf y},t_{\rm src},\dt,\dts) 
     +C_{3 \rm pt}^{\Gamma_{-}}({\bf y},t_{\rm src},N_t-\dt,N_t-\dts)
   \right\}
\label{3pt.average}
\end{eqnarray}
in order to reduce statistical errors.

\subsection{All-to-all quark propagator}
The three-point correlation function $C_{3\rm pt}$ is calculated 
by appropriately connecting the quark propagator $D^{-1}(x,y)$ 
as shown in Fig.~\ref{fig:diagram}.
The conventional method to calculate the quark propagator is not suitable 
to construct the disconnected quark loop starting from and ending at
arbitrary lattice sites 
since the source point $y$ has to be fixed at a certain lattice site.
Indeed, we use the all-to-all quark propagator technique,
which enables propagations 
from any lattice site to any site,
following the strategy proposed in \cite{Foley:2005ac,Bali:2005fu}. 

It is expected that 
low-lying eigenmodes of $D(m)$ dominantly
contribute to the low-energy dynamics of QCD.
We calculate the low-lying eigenvalues and eigenvectors
using the implicitly restarted Lanczos algorithm,
from which we can construct their contribution to the quark propagator
{\it exactly} as  
\begin{equation}
   (D^{-1}(m))_{\rm low}(x,y)
   =
   \sum_{i=1}^{N_e}\frac{1}{\lambda^{(i)}(m)}v^{(i)}(x)v^{(i)}(y)^{ \dagger},
   \label{DovL}
\end{equation}
where 
$\lambda^{(i)}(m)$ and $v^{(i)}(x)$ represent the $i$-th lowest eigenvalue 
and its associated eigenvector of $D(m)$, respectively.
Note that the eigenvectors are independent of valence quark masses.
The number of low-lying eigenmodes $N_e$ we calculated is 100 in this
study.

The remaining high-mode contribution is estimated stochastically. 
We prepare a single $Z_2$ noise vector $\eta(x)$ for each configuration
and split it into $N_{d}=3\times 4\times N_t/2$ vectors 
$\eta^{(d)}(x)$ ($d=1,...,N_d$), which have nonzero elements only
for a single combination of color and spinor indices 
on two consecutive time slices. 
The high-mode contribution is then estimated as
\begin{equation}
   (D^{-1}(m))_{\rm high}(x,y)
   =
   \sum_{d=1}^{N_{d}}\psi^{(d)}(x)\eta^{(d)}(y)^{\dagger} ,
   \label{DovH}
\end{equation}
where $\psi^{(d)}(x)$ is obtained by solving a linear equation for each
noise vector
\begin{equation}
 D(m) \psi^{(d)}(x) = (1-\mathcal P_{\rm low})\eta^{(d)}(x) \qquad (d=1,...,N_d).
\end{equation} 
$\mathcal P_{\rm low}$ is a projector 
to the subspace spanned by the low-modes
\begin{equation}
 \mathcal P_{\rm low}(x,y) = \sum_{i=1}^{N_e}v^{(i)}(x) v^{(i)}(y)^{\dagger}.
\end{equation}
We use this all-to-all propagator,
namely, (\ref{DovL}) plus (\ref{DovH}), 
to calculate the disconnected quark loop and the vacuum expectation
value of $S^{\rm lat}$ in $C_{3\rm pt}$.

\subsection{Low-mode averaging}
In principle, we can use the all-to-all propagator to calculate
nucleon correlators, namely, $C_{2\rm pt}$ and the piece representing the
nucleon propagation in $C_{3\rm pt}$. 
However, these quantities decay exponentially as the temporal
separation $\dt$ increases, so that the contributions to the nucleon
correlator from the high-modes (\ref{DovH}) are not sufficiently
precise at large $\dt$ when we take only one noise sample for each
configuration. 

In this study, we therefore use the low-mode averaging (LMA) technique
proposed in \cite{DeGrand:2004qw,Giusti:2004yp}.
Suppose that we decompose the conventional quark propagator 
into its low-mode part, which is in the subspace spanned by the low-modes
and the remaining high-mode part.
We can then write $C_{2\rm pt}$ in terms of the following eight contributions:
\begin{equation}
   C_{2\rm pt}
   = 
   C_{2\rm pt}^{lll}
  +C_{2\rm pt}^{llh}+C_{2\rm pt}^{lhl}+C_{2\rm pt}^{hll}
  +C_{2\rm pt}^{lhh}+C_{2\rm pt}^{hlh}+C_{2\rm pt}^{hhl}
  +C_{2\rm pt}^{hhh}.
  \label{C2pt:cntrb}
\end{equation}
Here, $C_{2\rm pt}^{lll}$ is constructed only by the low-mode part of the
quark propagator; $C_{2\rm pt}^{llh}$ is the one in which two of the valence
quarks are made of low-modes and the other is the high-mode part. 
The other combinations are understood in a similar manner. 
Since the ensemble average can be taken for each term of
(\ref{C2pt:cntrb}), we attempt to reduce the statistical error for
individual contributions. 

Relying on the translational invariance,
we may replace $C_{2\rm pt}^{lll}$ by a more precise estimate
by averaging over the location of the nucleon source point
$({\bf y},t_{\rm src})$.
No additional inversion of the Dirac operator is necessary to take the
average, as we can explicitly use the representation (\ref{DovL}) made
of low-mode eigenvectors. 
This LMA technique is very effective in reducing the statistical error 
of $C_{2\rm pt}$ at large $\dt$ when $C_{2\rm pt}$ is well dominated by
$C_{2\rm pt}^{lll}$. 

In this study, we employ LMA to calculate $C_{2\rm pt}$ and the nucleon
piece of $C_{3\rm pt}$. 
We also test an extension in which additional three contributions,
$C_{2\rm pt}^{llh}$, $C_{2\rm pt}^{lhl}$, and $C_{2\rm pt}^{hll}$, 
are averaged over the source location by using the all-to-all
propagator. 
The signal may be improved if the reduction of the statistical error
by the source average outweighs the induced noise from
the high-modes.
The result of the test is shown in the next section.

\subsection{Smeared nucleon operators}
Since $C_{2\rm pt}$ and $C_{3\rm pt}$ decay quickly as a function of $\dt$,
we need to use smeared nucleon operator that suppresses excited-state
contaminations at small $\dt$. 

For the (local or smeared) quark field, we consider the
following three choices: 
\begin{enumerate}
\item local
  \begin{equation}
    q_{\rm loc}({\bf x},t)=q({\bf x},t).
  \end{equation}
\item exponential smearing
  \begin{equation}
    q_{\rm smr}^{\rm exp}({\bf x},t) = \sum_{\bf r}\exp(-B|{\bf r} |) q({\bf {x+r}},t),
    \label{eqn:Exponential}
  \end{equation}
  where the parameter $B$ is set to 0.350, 0.375, 0.400 at
  $m_{ud}$ = 0.025, 0.035, 0.050, respectively.
\item Gaussian smearing
  \begin{equation}
    q_{\rm smr}^{\rm gss}({\bf x},t) 
    =
    \sum_{\bf y} \left\{ 
      \left( {\1}+\frac{\omega}{4N}H \right)^N 
    \right\}_{{\bf x,y}} 
    q({\bf y},t), \label{eqn:Gaussian} \qquad
    H_{{\bf x,y}} 
    =
    \sum_{i=1}^3 (\delta_{{\bf x,y}-\hat i}+\delta_{{\bf x,y}+\hat i}),
  \end{equation}
  where the parameters $\omega=20$ and $N=400$ are chosen so that 
  the extent of the smeared operator is roughly equal to that of 
  (\ref{eqn:Exponential}) with $B=0.400$. 
\end{enumerate}
Then, the nucleon interpolating fields $N_{\rm loc}({\bf x},t)$,
$N_{\rm smr}^{\rm exp}({\bf x},t)$, $N_{\rm smr}^{\rm gss}({\bf x},t)$,
are constructed from the corresponding local or smeared quark fields.

When we smear the quark field, we fix the gauge to the Coulomb gauge.
With this choice one can avoid significant statistical noise coming
from the fluctuation of the gauge link.

The Gaussian smearing is particularly useful for the sink smearing,
since the number of numerical operation~$\sim N\times N_s^3$ is smaller
than~$\sim N_s^6$ for the case of (\ref{eqn:Exponential}).
\section{Improving the nucleon two-point function}
\label{Sec:effective}
Since the disconnected three-point function $C_{3\rm pt}$ is extremely
noisy, it is crucial to reduce the statistical noise and to
extract the signal at relatively small time separations.
We therefore tested various methods to improve the signal on the
nucleon two-point functions $C_{2\rm pt}$ before applying them to the
three-point functions.

\subsection{Low-mode averaging}

\begin{figure}[tbp]
  \centering
  \includegraphics[width=0.49\textwidth,clip]{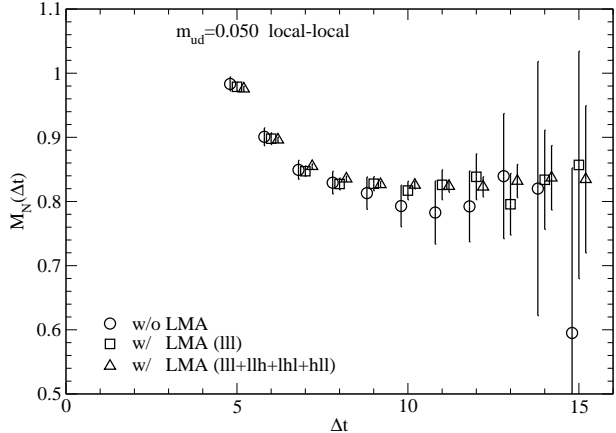}
  \caption{
    Effective mass $M_N(\dt)$ from the nucleon two-point function
    $C_{2\rm pt}$ at $m_{ud}=0.050$.
    The local operator is used for both source and sink.
    Circles show the result of the conventional point source, while
    squares (triangles) are obtained by averaging the 
    $C_{2\rm pt}^{lll}$~($C_{2\rm pt}^{lll}+C_{2\rm pt}^{llh}+C_{2\rm
 pt}^{lhl}+C_{2\rm pt}^{hll}$)
    contributions.
    Circles and triangles are slightly shifted in the horizontal
    direction for clarity.
  }
  \label{Fig:mN_lcl_m050}
\end{figure}

As mentioned in the previous section, we consider two options:
(i) to average only $C_{2\rm pt}^{lll}$ over the source locations,
(ii) to average also $C_{2\rm pt}^{llh}+C_{2\rm pt}^{lhl}+C_{2\rm pt}^{hll}$.
The second choice requires the high-mode of the quark propagator
$(D^{-1})_{\rm high}(x,y)$, which is calculated stochastically as in
(\ref{DovH}). 

In Fig.~\ref{Fig:mN_lcl_m050} we plot the nucleon effective mass
$M_N(\Delta t)$ with the local source and sink operators at our heaviest
quark mass $m_{ud}$ = 0.050.
The data without LMA (circles) show a rapidly growing statistical
error as $\Delta t$ increases, so that the error at $\Delta t$ = 10
where the plateau is approximately reached is already as large as 4\%.
By averaging over the source locations for $C_{2\rm pt}^{lll}$ (squares),
the statistical error is reduced by a factor of about 3.
Further improvement of a factor of 2 is possible if we average over
the source points also for
$C_{2\rm pt}^{llh}+C_{2\rm pt}^{lhl}+C_{2\rm pt}^{hll}$,
as shown by triangles.

\begin{figure}[tbp]
  \centering
  \includegraphics[width=0.49\textwidth,clip]{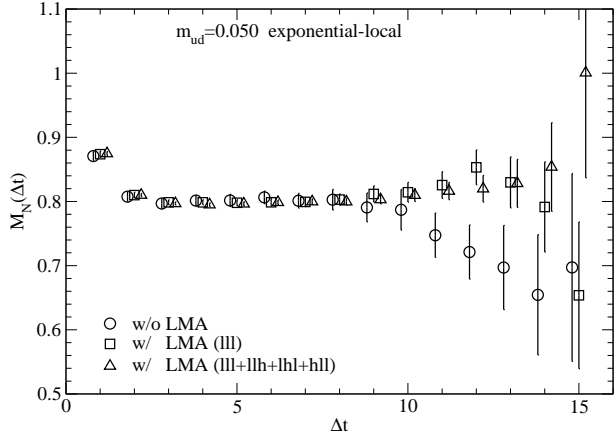}
  \caption{
    Effective mass $M_N(\dt)$ from $C_{2\rm pt}$ with an exponentially smeared
    source at $m_{ud}=0.050$.
    The symbols are the same as in Fig.~\protect\ref{Fig:mN_lcl_m050}.
 }
 \label{Fig:mN_exp_m050_Nsrc16}
\end{figure}

A similar comparison of $M_N(\Delta t)$ at $m_{ud}$ = 0.050 but with
the exponentially smeared source and a local sink is shown in
Fig.~\ref{Fig:mN_exp_m050_Nsrc16}. 
(But LMA is done over a limited number of the source location
$N_{\rm src}=N_t\times 16$. For discussions, see below.)
We observe that LMA for $C_{2\rm pt}^{lll}$ is efficient when combined
with the smeared source, while the effect of the extended LMA for
$C_{2\rm pt}^{llh}+C_{2\rm pt}^{lhl}+C_{2\rm pt}^{hll}$ is not substantial,
i.e., the reduction of statistical error is only about 30\%.

\begin{figure}[tbp]
  \centering
  \includegraphics[width=0.49\textwidth,clip]{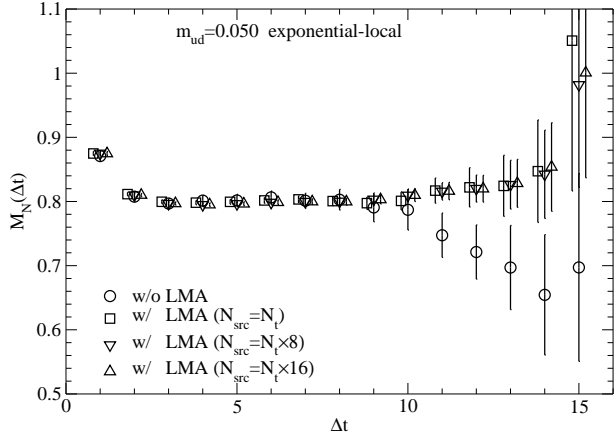}
  \caption{
    Comparison of $M_N(\dt)$ obtained with different numbers of source
    locations for LMA. 
    Circles are those without LMA.
    Results averaged over the time slices are shown by squares. 
    We obtain down- and up-triangles by further averaging over 8 and
    16 spatial sites at each time slice, respectively.
    In the plot, $N_{\rm src}$ represents the number of the source
    locations. 
  }
  \label{Fig:mN_exp_m050_Nsrc_cmp}
\end{figure}

Although the effect of LMA to reduce the statistical noise is
significant, it is also true that it requires substantial
computational effort.
If we average over the entire space-time source points, the
computational cost scales as $(N_s^3\times N_t)^2$, which is
prohibitive unless we use the fast Fourier transform.
If we combine LMA with the smeared source, another factor of
$N_s^3$ is necessary, which is not feasible any more.
We therefore consider averaging over a limited number of source
locations.
Since the correlators from different source points are statistically
highly correlated, this might not spoil the efficiency of LMA largely.

In Fig.~\ref{Fig:mN_exp_m050_Nsrc_cmp}, we compare the data of
$M_N(\Delta t)$ obtained using LMA 
with a different number of source points averaged $N_{\rm src}$.
The plot shows the results of LMA for both $C_{2\rm pt}^{lll}$ and
$C_{2\rm 2pt}^{llh}+C_{2\rm pt}^{lhl}+C_{2\rm pt}^{hll}$
with $N_{\rm src}$ = $N_t$ (squares), $N_t\times 8$ (triangles down),
and $N_t\times 16$ (triangles up).
For $N_{\rm src}=N_t$, the spatial location of the source is fixed
and the average is taken over $N_t$ time slices.
For $N_{\rm src}=N_t\times 8$, points of spatial coordinates 0 or
$N_s/2$ in three spatial dimensions are all averaged;
for $N_t\times 16$, we also average over
$(N_s/4,N_s/4,N_s/4)$, $(N_s/4,N_s/4,3N_s/4)$, 
$(N_s/4,3N_s/4,3N_s/4)$, and $(3N_s/4,3N_s/4,3N_s/4)$ (and all
possible permutations) for each time slice.

From Fig.~4 we observe that the result with $N_{\rm src}=N_t$ is
already very good, while the improvement with
$N_{\rm src}=N_t\times 8$ is not substantial.
Beyond this number, we do not gain significant improvement.
Note that the maximal number of points we took
$N_{\rm src}=N_t\times 16$ corresponds to the data shown in
Fig.~\ref{Fig:mN_exp_m050_Nsrc16} (triangles).

Overall, taking the cost of numerical calculation into account, the
best choice would be $N_{\rm src}\sim N_t\times 8$;
in our following analysis we choose $N_{\rm src}\sim N_t\times 16$,
which has been still doable.
The advantage of LMA for $C_{2\rm pt}^{lll}$ is always clear, while that for
$C_{2\rm pt}^{llh}+C_{2\rm pt}^{lhl}+C_{2\rm pt}^{hll}$ depends on the channel or
source operator.
Therefore, we average only $C_{2\rm pt}^{lll}$ when we use the smeared
sink, which is numerically more costly.

\subsection{Sink smearing}
The smearing of the source operator is routinely used in many lattice
calculations.
It is designed to deplete the overlap with excited-state contributions
so that the plateau of the effective mass constructed from the
two-point correlator appears earlier in $\Delta t$.
By using the smeared operator also for the sink we expect that the
excited-state contaminations are further reduced, but usually the
benefit is not clearly seen mainly because the statistical noise
increases with the smeared sink.
Since the numerical cost for the sink smearing is high in general
[$\sim(N_s^3)^2$], it has not been commonly used.

The situation may be different for three-point functions,
where an operator is inserted in the middle of the two-point function.
Here the nucleon and its excited states are created at the smeared
source point and propagate until the point of the operator is
reached.
Between these two points, the depletion of the excited states is at
work because of the smeared source.
After the insertion of the operator, the nucleon and its excited
states propagate until they are absorbed by the sink.
In this second propagation, the excited states are not necessarily
suppressed, since the operator insertion may excite the nucleon,
i.e., $\langle N|\bar{s}s|N'\rangle \neq 0$, and the sink operator
may have substantial overlap with the excited state $|N'\rangle$.
This is indeed the case in our calculation of the three-point function
relevant to the strange quark content, as we will see in the next
section.

\begin{figure}[tbp]
  \centering
  \includegraphics[width=0.49\textwidth,clip]{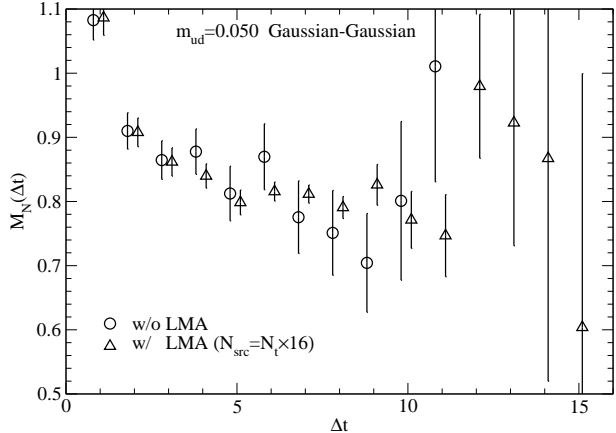}
  \caption{
    Effective mass $M_N(\dt)$ with the Gaussian smeared source and
    sink at $m_{ud}=0.050$. 
    Circle are without LMA;
    triangles are obtained by averaging $C_{2\rm pt}^{lll}$
    over 16 spatial sites at each time slice.
 }
 \label{Fig:mN_gss_m050_lll}
\end{figure}

We therefore utilize the smeared operator also for the sink.
Since the conventional choice $q_{\rm smr}^{\rm exp}(\mathbf{x},t)$
(\ref{eqn:Exponential}) 
requires a numerical cost proportional to $N_s^3$ for each
$(\mathbf{x},t)$, we use $q_{\rm smr}^{\rm gss}(\mathbf{x},t)$
(\ref{eqn:Gaussian}), instead. 
Figure~\ref{Fig:mN_gss_m050_lll} shows $M_N(\Delta t)$ with this
Gaussian smeared operator for both the source and sink.
Although the statistical signal is worse compared to the case of the
local sink shown in 
Fig.~\ref{Fig:mN_exp_m050_Nsrc16} and
\ref{Fig:mN_exp_m050_Nsrc_cmp}, 
we may improve it using LMA for $C_{2\rm pt}^{lll}$ as shown in
Fig.~\ref{Fig:mN_gss_m050_lll} by triangles.
Further improvement is not expected with the average over
$C_{2\rm pt}^{llh}+C_{2\rm pt}^{lhl}+C_{2\rm pt}^{hll}$,
as in the case of the smeared source and local sink
(Fig.~\ref{Fig:mN_exp_m050_Nsrc16}). 

\begin{figure}[tbp]
  \centering
  \includegraphics[width=0.49\textwidth,clip]{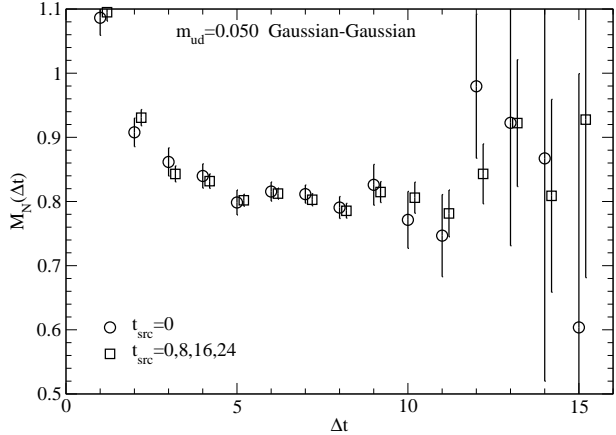}
  \caption{
    Improving the statistics by averaging the nucleon two-point
    functions calculated from four different source points at time
    slices $t_{\rm src}$ = 0, 8, 16, and 24.
    The result for $M_N(\dt)$ (squares) is compared with that without
    the average,  i.e., $t_{\rm src}$ = 0.
    The quark mass is $m_{ud}=0.050$.
    The Gaussian smearing is used for both the nucleon source and sink. 
 }
 \label{Fig:mN_gss_m050_Ntsrc}
\end{figure}

\subsection{Duplication}
Instead, we simply repeat the calculation 4 times by setting the
source at different time slices.
Namely, we calculate the nucleon two-point function
locating the source on the time slices $t_{\rm src}$ = 8, 16, and 24,
in addition to the original choice $t_{\rm src}=0$, and average over
these duplicated correlators.
The effect is shown in Fig.~\ref{Fig:mN_gss_m050_Ntsrc}, where we
observe a reduction of the statistical error by a factor of 2 at large
time separations. 
However, we find that the further average of the duplicated correlators
is not substantial. This is tested at $m_{ud}=0.025$ by 
calculating the nucleon two-point function locating the
source on the time slices $t_{\rm src}=4,12,20$, and $28$ 
besides $t_{\rm src}$ = 0, 8, 16, 24.
Therefore, we restrict the number of the duplication of the nucleon
two-point function for other quark masses.

\begin{figure}[tbp]
  \centering
  \includegraphics[width=0.49\textwidth,clip]{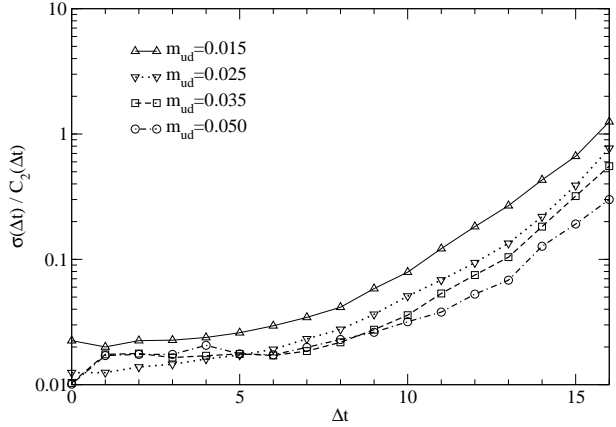}
  \caption{
    Noise-to-signal ratio of the nucleon correlator at $m_{ud}$ =
    0.015, 0.025, 0.035, and 0.050.
  }
  \label{Fig:sn_ratio}
\end{figure}

Figure~\ref{Fig:sn_ratio} shows the increase of the statistical noise in 
$C_{2\rm pt}$ for the case of smeared source and sink.
The plot shows the data at four different quark masses $m_{ud}$ =
0.015, 0.025, 0.035, and 0.050.
As expected, the noise grows more rapidly for lighter quarks.
Since the plateau in the effective mass is reached at around
$\Delta t=5$, we need at least $\Delta t=10$ in the calculation of the
three-point functions.
At the lightest quark mass $m_{ud}$ = 0.015, the error around 
$\Delta t=10$ is too large ($\sim 10\%$) to be useful in the analysis
of the disconnected three-point functions.
We therefore discard this data point in the analysis of the strange
quark content.

\begin{ruledtabular}
  \begin{table}[tbp]
    \begin{center}
      \begin{tabular}{l|| c c| c}
        Source-sink & $N_{\rm src}$ & LMAed contribution & Duplication \\
        \hline
        Local-local       & $N_t\times N_s$ &
        $C_{2\rm pt}^{lll}+C_{2\rm pt}^{llh}+C_{2\rm pt}^{lhl}+C_{2\rm pt}^{hll}$ & 1
        \\ 
        Exponential-local & $N_t\times 16$  &
        $C_{2\rm pt}^{lll}+C_{2\rm pt}^{llh}+C_{2\rm pt}^{lhl}+C_{2\rm pt}^{hll}$ & 1   
        \\
        Gaussian-Gaussian & $N_t\times 16$  & $C_{2\rm pt}^{lll}$       & 4
		   or 8 \\
      \end{tabular}
    \end{center}
    \caption{
      Choices of the scheme of averaging the nucleon correlator in
      this work.
      For different smearing operators at the source and sink,
      we list the number of source points $N_{\rm src}$ averaged in LMA, 
      the contributions to the correlator averaged in LMA
      ($C_{2\rm pt}^{lll}$ or 
      $C_{2\rm pt}^{lll}+C_{2\rm pt}^{llh}+C_{2\rm pt}^{lhl}+C_{2\rm pt}^{hll}$), and
      the number of the duplications of the conventional correlators.
    }
    \label{Tab:LMA}
  \end{table}
\end{ruledtabular}

In order to optimize the statistical signal in the calculation of the
disconnected three-point function for a given amount of computer time,
we choose different schemes of averaging the correlators depending on
the source and sink smearing combinations.
These include the choices of the contributions averaged in LMA 
($C_{2\rm pt}^{lll}$ or
$C_{2\rm pt}^{lll}+C_{2\rm pt}^{llh}+C_{2\rm pt}^{lhl}+C_{2\rm pt}^{hll}$),
the number of source points $N_{\rm src}$ averaged in LMA, as well as
the number of the duplications of the conventional correlators.
Our choices in this work are listed in Table~\ref{Tab:LMA}. 
\section{Extraction of the strange quark content}
\label{Sec:Results}

\subsection{Finding a plateau in the three-point function}
\begin{figure}[tbp]
  \centering
  \includegraphics[width=0.49\textwidth,clip]{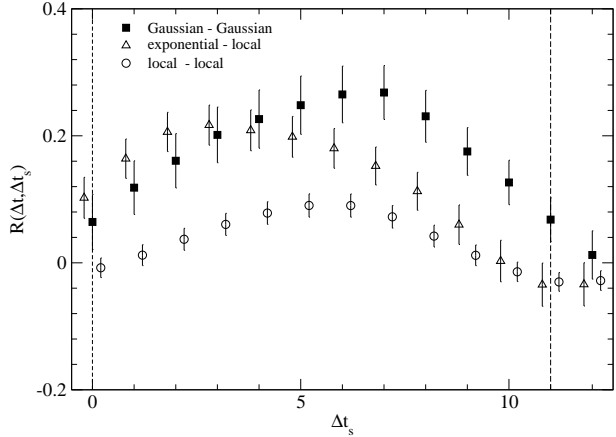}
  \caption{
    Ratio $R(\dt,\dts)$ with $\dt=11$ 
    at $m_{ud}=0.050$ and $m_{s,val}=0.100$. 
    Circles (triangles) are results obtained with the local
    (exponentially smeared) source and local sink, 
    whereas squares are calculated using the Gaussian smeared source
    and sink. 
    The vertical lines show the locations of the nucleon operators.
    The noisy high-mode contribution to the quark loop is ignored in
    this plot. 
  }
  \label{Fig:smr_m050_m100_dt11}
\end{figure}

We extract the strange quark content on the lattice
$\langle N|S^{\rm lat}|N\rangle$
from a ratio of 
$C_{3\rm pt}(\Delta t,\Delta t_s)$ and $C_{2\rm pt}(\Delta t)$
\begin{equation}
    R(\dt,\dts) 
   \equiv \frac{C_{3\rm pt}(\dt,\dts)}{C_{2\rm pt}(\dt)}
   \xrightarrow[\dt, \dts \to \infty\\ ]{ } 
   \langle N|S^{\rm lat}|N\rangle
   \label{Eqn:ratio}
\end{equation}
where $\Delta t$ is the temporal interval between the nucleon source
and sink.
The scalar operator $S^{\rm lat}$ is set on the time slice
apart from the nucleon source by $\Delta t_s$.
Note that $C_{3\rm pt}(\Delta t,\Delta t_s)$ and $C_{2\rm pt}(\Delta t)$
are calculated with LMA.
We suppress the coordinates of the nucleon source location
$(\mathbf{y},t_{\rm src})$ presented in (\ref{2pt.average}) and
(\ref{3pt.average}). 

In order to extract $\langle N|S^{\rm lat}|N\rangle$, we first have to
identify a plateau in the ratio $R(\Delta t,\Delta t_s)$ at sufficiently
large $\Delta t$ and $\Delta t_s$.
For this purpose, we look at the same ratio but approximated by taking 
only the low-mode contribution in the strange quark loop.
Namely, the piece of $S^{\rm lat}(z)$ in (\ref{3pt}) is replaced by its
low-mode contribution $\mathrm{Tr}[(D^{-1}(m))_{\rm low}(z,z)]$.
We expect that the ratio $R(\Delta t,\Delta t_s)$ is dominated by this
low-mode contribution, because the high-mode contribution that leads
to the ultraviolet divergence in the continuum limit cancels by the
VEV subtraction in (\ref{3pt}). 
Low-energy physics must be well described by the low-mode contribution
in the strange quark loop.
This approximation is finally removed in our calculation by the full
calculation, but here we consider the approximately calculated ratio
to identify the plateau, where the ground-state nucleon dominates.

\begin{figure}[tbp]
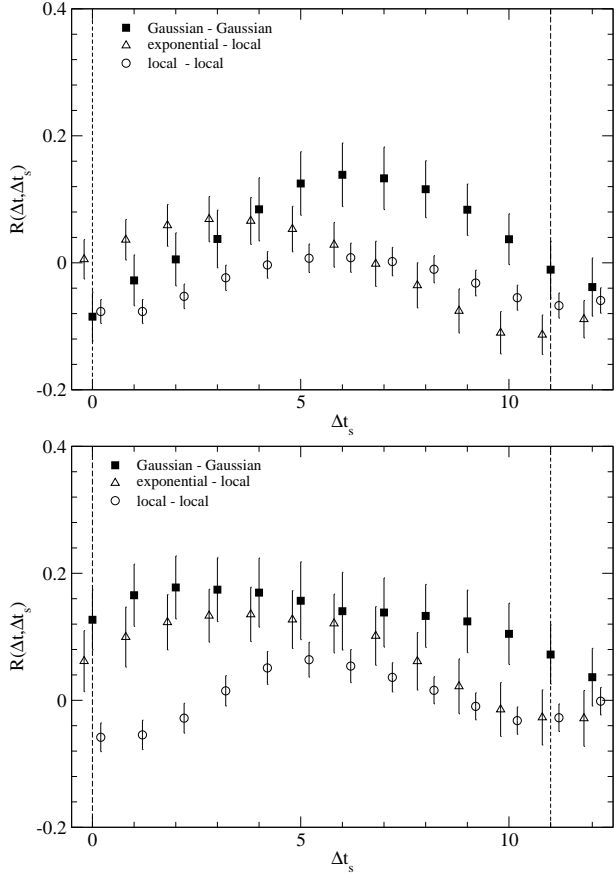

  \centering
  \includegraphics[width=0.49\textwidth,clip]{fig9.eps}
  \\
  \includegraphics[width=0.49\textwidth,clip]{fig10.eps}
  \caption{
    Same as Fig.~\ref{Fig:smr_m050_m100_dt11}, but for 
    $m_{ud}=0.035$ (top panel) and $m_{ud}=0.025$ (bottom panel).
  }
  \label{Fig:smr_m035_m100_dt11}
\end{figure}

Figure~\ref{Fig:smr_m050_m100_dt11} shows the approximated ratio
obtained at $m_{ud}=0.050$ and 
$m_{s,val}=0.100$ with various combinations of the source and sink
smearing. 
The separation between the source and sink is fixed to $\Delta t=11$,
and the location of the scalar operator $\Delta t_s$ is varied.
Thus, we expect a signal around $\Delta t_s \sim \Delta t/2$.
We observe a plateau between $\Delta t_s$ = 3 and 8, when the
source and sink operators are both smeared with the Gaussian smearing
(\ref{eqn:Gaussian}), as shown by filled squares. 
The data with the local source and sink (open circles) show a slight
increase in the same region but do not reach the value of the plateau
for the smeared source-sink combination.

The data of the smeared source and local sink (open triangles)
show a bump around $\Delta t_s\sim 2-6$ 
and decrease towards $\Delta t_s=11$, so that the plot looks
asymmetric. 
This can be explained by an excited-state contamination on the sink
side ($\Delta t_s=11$) because the sink operator is local.
Therefore, unlike the case for the two-point function, the use of the
smeared operator for both source and sink is essential for the
three-point function in order to extract the ground-state signal. 

Similar plots are shown for $m_{ud}$ = 0.035 and 0.025 in
Fig.~\ref{Fig:smr_m035_m100_dt11}.
We observe similar behavior of the approximated ratio.

\subsection{Bare results for the strange quark content}

\begin{figure}[tbp]
  \centering
  \includegraphics[width=0.49\textwidth,clip]{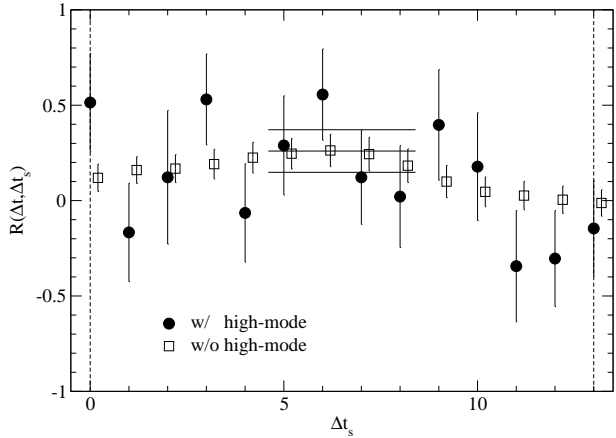}
  \caption{
    Ratio $R(\dt=13,\dts)$ at $m_{ud}=0.050$ and $m_{s,val}=0.100$
    with (filled circles) and without the high-mode contribution to
    the strange quark loop (open squares). 
    The horizontal lines show the result of a constant fit $R(\dt)$
    and its error band.
  }
  \label{Fig:fit_m050_m100_dt13}
\end{figure}

\begin{figure}[tbp]
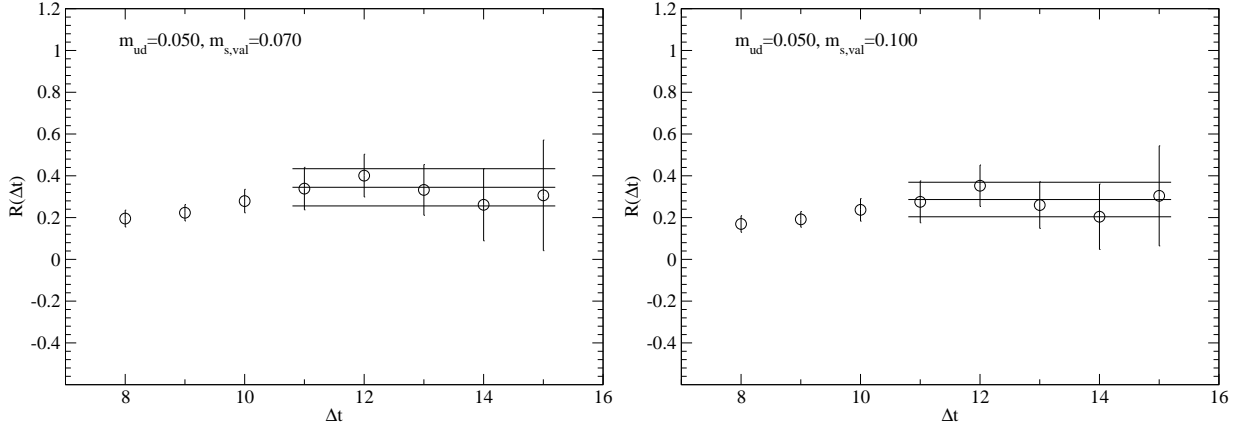

  \centering
  \includegraphics[width=0.49\textwidth,clip]{fig12.eps}
  \includegraphics[width=0.49\textwidth,clip]{fig13.eps}
  \caption{
    Results of the constant fit for $R(\dt,\dts)$ in the range
    $\dts=[5,\dt-5]$. 
    The data at $m_{ud}=0.050$. 
    The left and right panels show those at $m_{s,val}$ = 0.070 and
    0.100, respectively.
 }
 \label{Fig:dtfit_m050}
\end{figure}

The ratio $R(\Delta t,\Delta t_s)$ in (\ref{Eqn:ratio}) without the
low-mode approximation is shown in Fig.~\ref{Fig:fit_m050_m100_dt13}
(filled circles) together with that of the low-mode approximation
(open squares). 
Here, the data for $\Delta t$ = 13 are shown.
Although the statistical noise is much larger when the high-mode
contributions are included, the central value is unchanged.

Since the high-mode contributions are calculated with random noise
(\ref{DovH}), the larger noise is expected.
But, because the noise given for each time slice is statistically independent,
the correlation among the data points at different $\Delta t_s$ is
expected to come mainly from the low modes,
provided that the high-mode contribution to the ratio is negligible,
which is indeed the case within our statistical accuracy. 
The statistical error is then effectively reduced by averaging over
different $\Delta t_s$. 
In Fig.~\ref{Fig:fit_m050_m100_dt13}, the result of a constant fit for
$\Delta t_s$ = [5,8] is 
shown by a horizontal line together with a band showing the resulting
statistical error.
In this case, the statistical error of the fitted value is about a
half of that of each point, because four data points are averaged.
We also checked that the statistical correlation among the points at
different $\Delta t_s$ is an order of magnitude smaller than the
variance of each point.

For the final result, we take the full data including the high modes
and fit in the region where the approximated ratio shows a plateau.
To be specific, we fit in the region $\Delta t_s=[5,\Delta t-5]$ with
$\Delta t \ge 11$.

\begin{figure}[tbp]
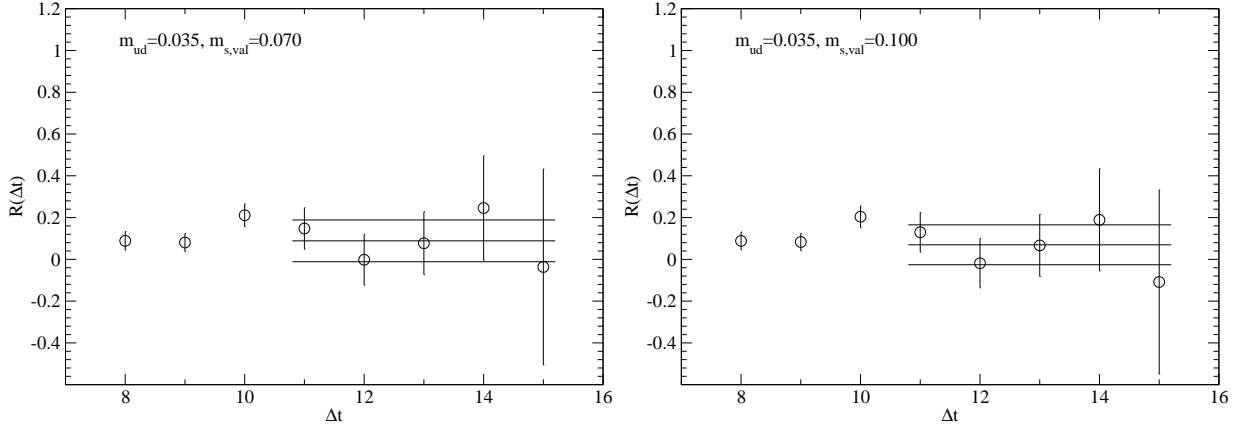

  \centering
  \includegraphics[width=0.49\textwidth,clip]{fig14.eps}
  \includegraphics[width=0.49\textwidth,clip]{fig15.eps}
  \caption{
    Same as Fig.~\ref{Fig:dtfit_m050} but at $m_{ud}=0.035$.
 }
 \label{Fig:dtfit_m035}
\end{figure}

\begin{figure}[tbp]
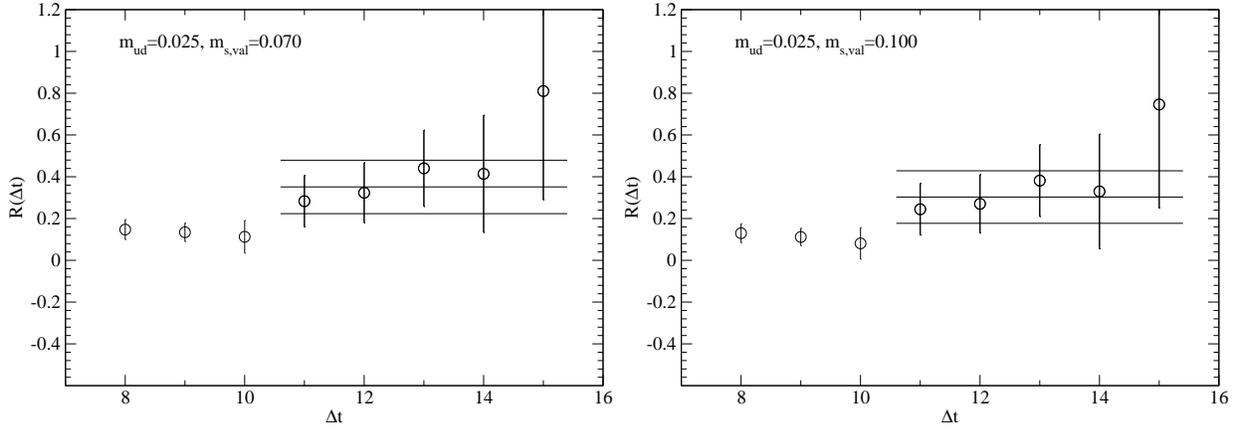

  \centering
  \includegraphics[width=0.49\textwidth,clip]{fig16.eps}
  \includegraphics[width=0.49\textwidth,clip]{fig17.eps}
  \caption{
    Same as Fig.~\ref{Fig:dtfit_m050} but at $m_{ud}=0.025$.
  }
  \label{Fig:dtfit_m025}
\end{figure}

Figures~\ref{Fig:dtfit_m050}--\ref{Fig:dtfit_m025} show the results of
the constant fit for each $\Delta t$. 
We find that the results are stable under the change of $\Delta t$.
We then fit these results by a constant in $\Delta t$ = [11,15].
The statistical error is estimated using the jackknife method.
The numerical results are listed in Table~\ref{Tab:Bare values}.

In order to estimate the systematic effect due to possible contamination
of the excited states, we also test a fitting form for $R(\dt,\dts)$
taking account of the first excited state:
\begin{equation}
 R(\dt, \dts) 
= 
c_0 - c_1 e^{-(2M_0 + \Delta M)\Delta t/2} 
           \cosh(\Delta M (\Delta t_s - \Delta t/2)), 
\end{equation}
where the first and second terms represent the contributions
from the ground and first excited states, respectively. $\Delta M$ is
the mass gap between these two states.  
To make this fit stable, we carry out a simultaneous fit 
in terms of $\dts$ and $\dt$ using a slightly wider 
fit range, $\dts=[4,\Delta t-4]$ and $\dt \geq 11$.
We also use the ground-state mass $M_0$ determined from the 
nucleon two-point function.
The excited-state contribution represented by the $c_1$ term turned out to be
small: in the maximum case ($\Delta t=11$) it is about 0.04(8) compared
to the main contribution $c_0\simeq $ 0.3(1). For large $\Delta t$, the
excited-state contribution is more suppressed.
This is expected from the small
$\dts$ and $\dt$ dependence of the ratio shown in
Figs.\ref{Fig:fit_m050_m100_dt13}--\ref{Fig:dtfit_m025}. We 
therefore use the results in Table~\ref{Tab:Bare values} in the following
analysis without adding further errors due to the excited states.

\begin{ruledtabular}
  \begin{table}[tbp]
    \begin{center}
      \begin{tabular}{cccc}
        $m_{ud}$ & Fit range of $\dt$ 
        & $m_{s,val}=0.070$ & $m_{s,val}=0.100$ \\
        \hline
        0.050    &   [11,15]   & 0.345(89)  & 0.286(83) \\
        0.035    &   [11,15]   & 0.089(100) & 0.070(96) \\
        0.025    &   [11,15]   & 0.351(128) & 0.303(126)\\
      \end{tabular}
    \end{center}
    \caption{
      Strange quark content $\langle N|S^{\rm lat}|N\rangle$
      calculated on the lattice at each quark mass. 
      The fit range of $\dt$ is also listed.
    }
    \label{Tab:Bare values}
  \end{table}
\end{ruledtabular}

\section{Chiral extrapolation to the physical point}
\label{Sec:Extrap.}

In this section, we discuss on the extrapolation of our lattice data
to the physical quark masses.
We have three data points corresponding to up and down quark masses $m_{ud}$ in 
the range of $M_\pi$ = 370--520~MeV.
For the strange quark mass we have two data points sandwiching the
physical strange quark mass.

\begin{figure}[tbp]
  \centering
  \includegraphics[width=0.49\textwidth,clip]{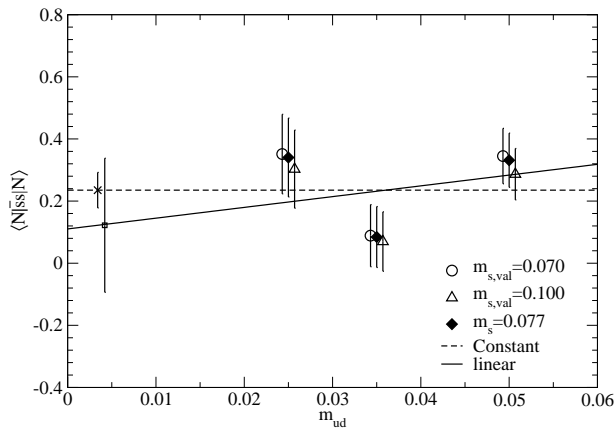}
  \caption{
    The dependence of $\langle N|S^{\rm lat}|N\rangle$ on the up and down
    quark mass $m_{ud}$ (given in the lattice unit).
    Open circles and triangles are the data at each $m_{ud}$ and $m_s$
    = 0.070 (circles) and 0.100 (triangles).
    The data linearly interpolated to the physical strange quark mass
    $m_{s,\rm phys}$ is shown by filled diamonds.
    Dashed and solid lines show the fit curve at $m_{s,\rm phys}$
    obtained from the constant and linear extrapolations.
  }
 \label{Fig:chextrap_linear}
\end{figure}

Our data for the matrix element $\langle N|S^{\rm lat}|N\rangle$ are
plotted as a function of $m_{ud}$ in Fig.~\ref{Fig:chextrap_linear}.
We do not observe statistically significant dependence of 
$\langle N|S^{\rm  lat}|N\rangle$ on both $m_{ud}$ and $m_s$.
By fitting the data linearly in $m_{ud}$ and $m_s$ as
\begin{equation}
   \langle N|S^{\rm  lat}|N\rangle
   = c_0+c_{1,ud}m_{ud}+c_{1,s}m_{s,val},
   \label{eqn:linear}
\end{equation}
we obtain the numerical results of the fit parameters $c_0$,
$c_{1,ud}$, and $c_{1,s}$ listed in Table~\ref{Tab:extrap.}.
We also show the result of a constant fit including only the $c_0$
term in (\ref{eqn:linear}).
Both results are consistent with each other, but the linear
extrapolation gives a larger error at the physical point.

\begin{ruledtabular}
  \begin{table}[tbp]
    \begin{center}
      \begin{tabular}{l|cclllc}
        &
        $\chi^2/{\rm d.o.f.}$ & {\rm d.o.f.} & 
        $c_0$ & $c_{1,ud}$ & $c_{1,s}$ & 
        $\langle N|S^{\rm lat}|N\rangle$
        \\ \hline
        constant & 
        1.63  &  5 & 0.24(6)  & $\cdots$          & $\cdots$          & 0.24(6) 
        \\
        linear & 
        2.39  &  3 & 0.22(24) & 3.5(5.7) & $-$1.44(52) & 0.12(22)
        \\
      \end{tabular}
    \end{center}
    \caption{
      Numerical results of chiral extrapolation.
      We also list $\langle N|S^{\rm lat}|N\rangle$ extrapolated to
      the physical point. 
    }
    \label{Tab:extrap.}
  \end{table}
\end{ruledtabular}

Assuming that the quark mass dependence of the nucleon mass is
reliably described by the chiral perturbation theory, we also
attempt an extrapolation using the formula provided by the
$SU(3)$ heavy baryon chiral perturbation theory (HBChPT).
From the chiral expansion of $M_N$~\cite{WalkerLoud:2004hf} and the
Feynman-Hellmann theorem~(\ref{eqn:FH}), which will be discussed in
Sec.~\ref{Sec:Renorm.}, 
the quark mass dependence of $\langle N|S^{\rm lat}|N\rangle$
up to the next-to-leading order is given by 
\begin{equation}
  \langle N|S^{\rm lat}|N\rangle
  = -c_s -B
  \left\{ \frac{3}{2} C_{NNK} \, M_K + 2 C_{NN\eta}\, M_\eta \right\},
  \label{eqn:NLO}
\end{equation}
where the coefficients $C_{NNK}$ and $C_{NN\eta}$ are written as
\begin{eqnarray}
  C_{NNK} &=& \frac{1}{8\pi f^2}\frac{(5D^2-6DF+9F^2)}{3},
  \label{eqn:NLO:coeff1}
  \\
  C_{NN\eta} &=& \frac{1}{8\pi f^2}\frac{(D-3F)^2}{6}.
  \label{eqn:NLO:coeff2}
\end{eqnarray}
The axial couplings $F$ and $D$ are phenomenologically well determined 
and we fix them as $D=0.81$ and $F=0.47$ \cite{Jenkins:1991es}.
For the pseudoscalar meson masses $M_K$ and $M_\eta$, we use the
Gell-Mann, Oakes, and Renner (GMOR) relations
$M_K^2 = B(m_{ud}+m_s)$ and $M_\eta^2 = 2 B(m_{ud}+2 m_s)/3$,
which are valid at the leading order of the quark masses.
We fix the low-energy constants $f$ and $B$
to the values obtained in our study of the pion mass
and decay constant \cite{Noaki:2008iy}.
Note that the contributions of the decuplet baryons are ignored in
this analysis. 

\begin{figure}[tbp]
  \centering
  \includegraphics[width=0.49\textwidth,clip]{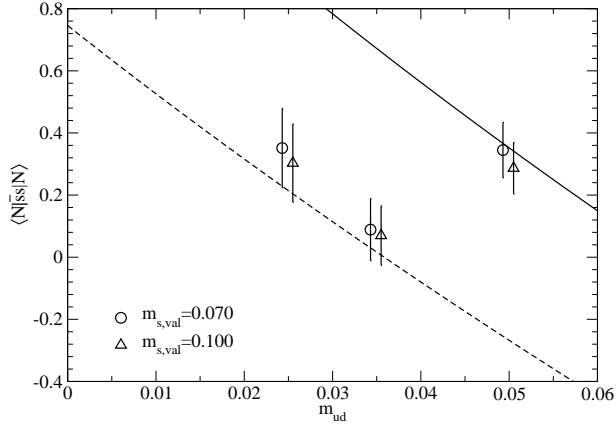}
  \caption{
    The chiral fit of $\langle N|S^{\rm lat}|N\rangle$
    based on the next-to-leading order HBChPT~(\ref{eqn:NLO}).
    Solid and dashed lines show the fits 
    at $m_{s,val}=0.070$ and $0.100$. 
 }
 \label{Fig:phenom_nlo}
\end{figure}

As one can see from Fig.~\ref{Fig:phenom_nlo}, this function does
not describe the numerical data; the value of $\chi^2$ per degree of
freedom (d.o.f.) is unacceptable ($\sim 20$).
The main reason is that there is no free parameter to control the
quark mass dependence, i.e., the coefficients of $M_K$ and
$M_\eta$ in (\ref{eqn:NLO}) are completely determined
phenomenologically.
In other words, if we leave $f$ as a free parameter for instance, the
resulting value is unreasonably large.

\begin{figure}[tbp]
  \centering
  \includegraphics[width=0.49\textwidth,clip]{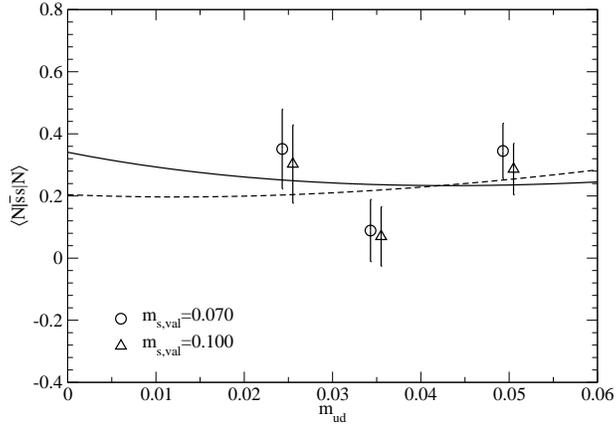}
  \caption{
    The chiral fit using (\ref{eqn:NNLO}) with a higher order term.
  }
  \label{Fig:phenom_nnlo}
\end{figure}

If we add a higher order analytic term as
\begin{equation}
  \langle N|S^{\rm lat}|N\rangle
   = -c_s -B
   \left\{ \frac{3}{2} C_{NNK} \, M_K + 2 C_{NN\eta}\, M_\eta \right\}
   +c_2 M_K^2,
   \label{eqn:NNLO}
\end{equation}
the fit becomes reasonable as shown in Fig.~\ref{Fig:phenom_nnlo},
for which $\chi^2/{\rm d.o.f.}$ is acceptable ($\sim$ 1.9).
Fit parameters obtained with (\ref{eqn:NLO}) and (\ref{eqn:NNLO}) are
summarized in Table~\ref{Tab:phenom.}.
The resulting fit parameters suggest that the chiral expansion
does not converge well.
In fact, if we look at the individual contributions to 
$\langle N|S^{\rm lat}|N\rangle$ from each term in (\ref{eqn:NNLO}),
all of them are an order of magnitude larger than the
data themselves, and the final result is obtained by a large
cancellation. 

\begin{ruledtabular}
  \begin{table}[tbp]
    \begin{center}
      \begin{tabular}{l|ccccc}
        &
        $\chi^2/{\rm d.o.f.}$ & ${\rm d.o.f}$ &
        $-c_s$ & $c_2$ & $\langle N|S^{\rm lat}|N\rangle$
        \\ \hline
        Equation~(\ref{eqn:NLO})  &  19.5 & 5 & 5.48(6) & $\cdots$ & 1.24(6)\\
        Equation~(\ref{eqn:NNLO}) &  1.88 & 4 & 2.82(23) & 21.2(1.8) & 0.28(10)
      \end{tabular}
    \end{center}
    \caption{
      Numerical results of chiral fits using the $SU(3)$ HBChPT formulas, 
      i.e., (\ref{eqn:NLO}) and (\ref{eqn:NNLO}).
    }
    \label{Tab:phenom.}
  \end{table}
\end{ruledtabular}

Because of this poor convergence of the chiral expansion, we use the
result of the HBChPT analysis only to estimate the systematic
uncertainty. 
Namely, we take the result from the constant fit as a central value of 
$\langle N|S^{\rm lat}|N\rangle$ at the physical quark masses.
The systematic error due to the chiral extrapolation 
is estimated by a difference from the results of the 
linear (\ref{eqn:linear}) and HBChPT fits (\ref{eqn:NNLO}).
Then, we obtain
$\langle N|S^{\rm lat}|N\rangle = 0.24(6)(16)$
at the physical quark masses.
The first and second errors represent the statistical and systematic ones.

Using the experimental value of $M_N$, this is converted to the
strange quark mass contribution to $M_N$ defined in (\ref{eqn:fts}) as
\begin{equation}
  f_{T_s} = 0.032(8)(22).
  \label{eqn:fTs_result}
\end{equation}
Since the combination $m_sS^{\rm lat}$ is invariant under
renormalization, no renormalization factor is required to obtain
(\ref{eqn:fTs_result}). 

The $y$ parameter (\ref{eqn:y}) is defined as a ratio of the strange
and {\it ud} quark contents. 
We obtain 
\begin{equation}
  y = 0.050(12)(34),
  \label{eqn:y_result}
\end{equation}
where we use an estimate 
$\langle N|\bar uu + \bar dd |N \rangle \!=\! 9.40(41)$ 
for the denominator, which is taken from our study
of the nucleon sigma term \cite{Ohki:2008ff}.

A simple order counting suggests that the discretization effect is
$O((a \Lambda)^2) \sim$ 9\% when we take $\Lambda\sim$ 500~MeV.
Other systematic errors including those of finite volume effects would
not be significant, given that the statistical and systematic errors in
(\ref{eqn:fTs_result}) and (\ref{eqn:y_result}) are so large 
($\sim$ 70\%).

\section{Comparison with previous lattice calculations}
\label{Sec:Renorm.}

In this section, 
we emphasize an important role played by the exact chiral symmetry
in the calculation of the strange quark content.
Then we compare our result with the previous calculations.

\subsection{Renormalization issue of the operator $\bar{s}s$}
First, let us consider the renormalization of the $\bar{s}s$ operator
in the flavor $SU(3)$ symmetric limit for simplicity.
Using the flavor triplet quark field $\psi$, the $\bar{s}s$ operator
can be written in terms of flavor-singlet and octet operators as  
\begin{equation}
  (\bar{s}s)^{\rm phys} 
  = 
  \frac{1}{3}
  \left\{
   (\scalar)^{\rm phys} 
   -\sqrt{3}\, (\bar{\psi}\lambda^8 \psi)^{\rm phys}
  \right\},
\label{eqn:singlet_octet}
\end{equation}
where $\lambda^8$ is a Gell-Mann matrix.
Note that, in this section, 
we put the superscript ``phys'' on the renormalized quantities
defined in the continuum theory to distinguish them from bare
operators, which is in our case defined on the lattice. 

\begin{figure}[t]
 \centering
 \includegraphics[width=0.4\textwidth,clip]{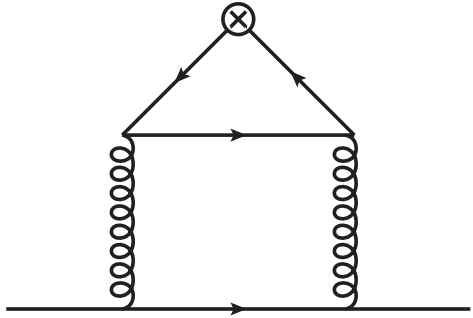}
 \caption{
   The disconnected diagram contributing to the renormalization of
   flavor-singlet scalar operator (cross). 
   At higher orders, the quark loop and the quark line on the bottom
   may be connected by an arbitrary number of gluon propagators.
   Since the quark-quark-gluon vertex conserves chirality, the
   chirality of the quark propagating in the loop does not change, as far
   as the regularization respects chiral symmetry.
 }
 \label{Fig:sigl.daig}
\end{figure}

In general, the singlet and octet operators may be renormalized differently
\begin{eqnarray}
 (\scalar)^{\rm phys}
  &=& 
  Z_0\, (\scalar)
  \label{singlet},\\
 ( \bar{\psi}\lambda^8 \psi)^{\rm phys}
  &=& 
  Z_8\, ( \bar{\psi}\lambda^8 \psi),
  \label{octet}
\end{eqnarray}
with different renormalization factors $Z_0$ and $Z_8$.
Here, we assume that the chiral symmetry is preserved in the
renormalization scheme used to calculate (\ref{singlet}) and (\ref{octet}).
Otherwise, there is a mixing with lower dimensional operators
for the flavor-singlet operator (\ref{singlet}), as discussed below.
The operator $(\bar{s}s)^{\rm phys}$ is then expressed in terms of bare
operators as  
\begin{equation}
   (\bar{s}s)^{\rm phys}  =
   \frac{1}{3}
   \left\{
      (Z_0 + 2 Z_8) (\bar{s}s)
     +(Z_0 - Z_8)(\bar{u}u+\bar{d}d)
   \right\},
   \label{eqn:sbars:gen}
\end{equation}
which implies that the $\bar{s}s$ can mix with $\bar{u}u+\bar{d}d$
unless $Z_0=Z_8$.
The difference $Z_0-Z_8$ arises from disconnected diagrams such as
those shown in Fig.~\ref{Fig:sigl.daig}, which exist only for the
flavor-singlet operator.

When the renormalization scheme respects chiral symmetry,
the disconnected diagrams vanish in the massless limit, because 
the quark loop starting from and ending at a scalar operator 
$\bar{s}s=\bar{s}_Ls_R+\bar{s}_Rs_L$
has to change the chirality in the loop while the change of chirality
does not occur by attaching any number of gluon lines to the quark
loop. 
It means that $Z_0=Z_8$ is satisfied for mass independent
renormalization schemes, as far as they maintain exact chiral
symmetry.
This also applies in the case of the overlap fermion formulation on
the lattice, as there is an exact chiral symmetry guaranteed by the
Ginsparg-Wilson relation \cite{Ginsparg:1981bj} at finite lattice
spacings \cite{Luscher:1998pqa}.

Thus, the renormalization of the scalar operator reduces to a
multiplicative renormalization 
$(\bar{s}s)^{\rm phys}(\mu) = Z_S(\mu) S^{\rm lat}$
with $Z_S=Z_0=Z_8$.
Here we specify the renormalization point $\mu$ for the renormalized
operator $(\bar{s}s)^{\rm phys}$.
The value of $Z_S(\mu)$ is nonperturbatively calculated in
\cite{Noaki:2009xi} as $Z_S(\mathrm{2~GeV})$ = 1.243(15) on our
lattice. 
For the numerical results of $f_{T_s}$ (\ref{eqn:fTs_result}) and $y$
(\ref{eqn:y_result}) quoted in the previous
section, the renormalization factor is unnecessary, because they are
related to a renormalization invariant operator $m_s\bar{s}s$ or a
ratio $\bar{s}s/(\bar{u}u+\bar{d}d)$.

As it is clear from the above discussion,
the explicit violation of chiral symmetry with the conventional
Wilson-type fermions induces a mixing between the strange and $ud$ 
quark contents.  
In addition, the flavor-singlet scalar operator mixes with an identity 
operator, so that (\ref{eqn:sbars:gen}) is modified as 
\begin{equation}
   (\bar{s}s)^{\rm phys} 
   = 
   \frac{1}{3}
   \left[
      (Z_0 + 2 Z_8) (\bar{s}s) 
      +(Z_0 - Z_8)(\bar{u}u+\bar{d}d)
     + \frac{b_0}{a^3} + \cdots
   \right],
\end{equation}
where the term $b_0/a^3$ represents the power divergent mixing
contribution. 
This contribution from the identity operator must be subtracted 
as a part of the vacuum expectation value of $\bar{s}s$.
Because of the cubic divergence, this results in a large cancellation
toward the continuum limit.

Furthermore, since $Z_0-Z_8$ does not vanish when chiral symmetry is
violated, $\bar{s}s$ mixes with
$\bar{u}u+\bar{d}d$, which induces a connected diagram contribution in
the calculation of the three-point function. 
Since the connected diagram is larger than the disconnected
contribution by an order of magnitude, the whole effect from 
$(Z_0-Z_8)(\bar{u}u+\bar{d}d)$ could be substantial, even though the
difference $Z_0-Z_8$ may be small.
This possibility has been neglected in most of the previous lattice
calculations using the Wilson-type fermions.

\subsection{Direct and indirect calculations}
The strange quark content can also be calculated 
from the $m_s$ dependence of $M_N$ through the Feynman-Hellmann theorem
\begin{equation}
 \langle N | \bar{s}s | N \rangle
  = 
  \frac{\partial M_N}{\partial m_{s}}.
  \label{eqn:FH}
\end{equation}
We refer to this method as the spectrum method in the following.
Exact chiral symmetry plays a crucial role in this method, too.
With the explicit chiral symmetry violation,
masses of sea and valence quarks, $m_{f,\rm sea}$ and $m_{f,\rm val}$ (where $f$
distinguishes the quark flavors $ud$ and $s$),
depend on the sea strange quark mass $m_{s,\rm sea}$. 
Namely, there is an additive mass renormalization $\Delta m$
\begin{eqnarray}
  m_{f,\rm sea}^{\rm phys} &=& Z_m(m_{f, \rm sea} +\Delta m), \\
  m_{f,\rm val}^{\rm phys} &=& Z_m(m_{f, \rm val} +\Delta m), 
\end{eqnarray}
when we relate the bare quark masses on the lattice ($m_{f,\rm sea}$ and
$m_{f,\rm val}$) to their counterparts ($m_{f,\rm sea}^{\rm phys}$ and
$m_{f,\rm val}^{\rm phys}$) defined in some continuum renormalization scheme.
$Z_m$ is the multiplicative renormalization factor.
With dynamical Wilson fermions, this additive mass renormalization
$\Delta m$ is of the cutoff order, $\sim 1/a$, and its dependence on
the sea quark mass is a quantity of order unity.

Then, we can write the relevant partial derivative 
$\partial M_N/\partial m_{s,\rm sea}$
calculated on the lattice in terms of the ``physical'' quark mass
dependence of $M_N$ as
\begin{eqnarray} 
   \frac{\rd M_N}{\rd m_{s,\rm sea}} 
   & = &
   \frac{\rd m_{s,\rm sea}^{\rm phys}}{\rd m_{s,\rm sea}}
   \frac{\rd M_N}{\rd m_{s,\rm sea}^{\rm phys}}
   +
   \frac{\rd m_{ud,\rm sea}^{\rm phys}}{\rd m_{s,\rm sea}}
   \frac{\rd M_N}{\rd m_{ud,\rm sea}^{\rm phys}}
   +
   \frac{\rd m_{ud,\rm val}^{\rm phys}}{\rd m_{s,\rm sea}}
   \frac{\rd M_N}{\rd m_{ud,\rm val}^{\rm phys}} 
\notag \\
   & = &
   Z_m \left[
          \NssN^{\rm phys}
         +\frac{\rd \Delta m}{\rd m_{s,\rm sea}}
          \langle N|\bar{u}u+\bar{d}d+\bar{s}s|N \rangle^{\rm phys}
       \right],
  \label{eqn:prtl_drvtv}
\end{eqnarray}
where the matrix elements appearing on the right-hand side are those
with the continuum renormalization scheme.
The last term must be subtracted from 
$(1/Z_m) \partial M_N/\partial m_{s,\rm sea}$
to obtain the strange quark content.
It requires a calculation of the light quark content
$\langle N|\bar{u}u+\bar{d}d|N\rangle$, which is dominated by the
connected diagram, and of the $m_{s,\rm sea}$ dependence of $\Delta m$,
which strongly depends on the details of the lattice action used in
the calculation.
In the literature, this subtraction was considered only in
\cite{Michael:2001bv}, where the subtraction induced a rather large
statistical error.

One may avoid this problem by differentiating $M_N$ in terms of pion
and kaon mass squared, $M_\pi^2$ and $M_K^2$, instead of $m_s$,
assuming the GMOR relations $M_\pi^2=2Bm_{ud}$, $M_K^2=B(m_{ud}+m_s)$.
Since the quark masses appearing in the right-hand side of the GMOR
relations contain the additive mass renormalization $\Delta m$, 
the above subtraction is not necessary.
But the method introduces another uncertainty, because the GMOR
relations are valid only at the leading order of the quark mass, and
the higher order terms are not negligible in general.
This method has been applied in the analysis of \cite{Young:2009zb}.

\begin{figure}[tbp]
  \centering
  \includegraphics[width=.63\textwidth,clip]{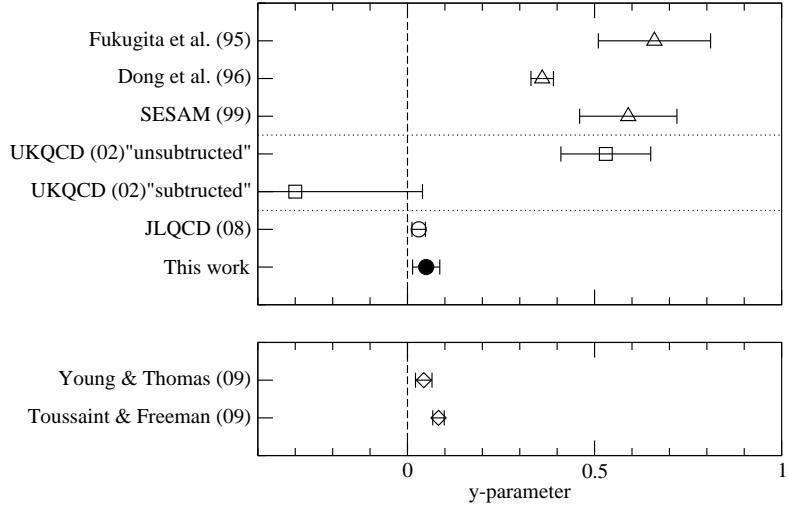}
   \vspace{-2mm}
  \caption{
    (Top panel)~The comparison of the $y$ parameter with previous studies.
    The result in this study (\ref{eqn:y_result}) and 
    our estimate using the spectrum method \cite{Ohki:2008ff} 
    are plotted by filled and open circles.
    Triangles represent previous direct calculations from the nucleon
    three-point functions  
    \cite{Fukugita:1994ba,Dong:1995ec,Gusken:1998wy}. 
    Squares are the results of \cite{Michael:2001bv} that used the
    spectrum method. 
    We plot two results obtained with and without subtracting the
    contamination due to the chiral symmetry breaking.
    (Bottom panel)~Two recent results
    \cite{Young:2009zb,Toussaint:2009pz} are plotted. 
    Values are converted to the $y$ parameter using the quark mass ratio 
    $m_s/m_{ud}=27.4(4)$~\cite{Aubin:2004fs} and the nucleon $\sigma$
    term \cite{Ohki:2008ff}. 
    The quoted errors are statistical only except for our studies and
    the two recent calculations \cite{Young:2009zb,Toussaint:2009pz}.
  }
  \label{fig:comp}
\end{figure}

In Fig.~\ref{fig:comp} (top panel) we compare our result
(\ref{eqn:y_result}) for the $y$ parameter plotted by a solid circle
with those from previous studies using the Wilson-type actions
\cite{Fukugita:1994ba,Dong:1995ec,Gusken:1998wy,Michael:2001bv}.
Among these, \cite{Fukugita:1994ba,Dong:1995ec} are quenched
calculations and \cite{Gusken:1998wy,Michael:2001bv} contain the
effects of two dynamical flavors.
Rather large values $y$ = 0.4--0.8 were obtained in the calculations 
from the nucleon three-point functions 
\cite{Fukugita:1994ba,Dong:1995ec,Gusken:1998wy}, 
for which the above mentioned contamination was not taken into
account and large systematic error is expected.
An exception is the UKQCD's calculation with the spectrum method
\cite{Michael:2001bv}; 
the subtraction of the contamination led to a large uncertainty in $y$.

In the same figure, we also compare our result (\ref{eqn:y_result}) 
with our previous estimate $y = 0.030(16)(^{+6}_{-8})$
from the spectrum method \cite{Ohki:2008ff},
where the first and second errors are statistical and systematic,
respectively. 
Because of the exact chiral symmetry satisfied in both of our
calculations, these two points are free from the contamination and
consistent with each other.

Recently there have been two calculations published
\cite{Young:2009zb,Toussaint:2009pz}.
The analysis of Young and Thomas \cite{Young:2009zb} fits the data
from recent calculations of the baryon spectrum done by the LHPC
\cite{WalkerLoud:2008bp} and PACS-CS \cite{Aoki:2008sm}
Collaborations, and takes a derivative in terms of $M_\pi^2$ and 
$M_K^2$.
As already mentioned, the problem of the operator mixing is avoided in
this method and the authors obtained a result consistent with ours.
Toussaint and Freeman \cite{Toussaint:2009pz} uses the data for the 
nucleon mass obtained by the MILC Collaboration using the so-called
``asqtad'' quark action, which is a variant of the staggered fermion
formulation.
They use a clever idea of extracting the derivative of the nucleon
correlator in terms of the quark mass from the correlation between the
nucleon correlator and the scalar density operator (the
Feynman-Hellmann theorem).
Since the staggered fermion has a remnant chiral symmetry, there is no
problem with the operator mixing.
On the other hand, there is a subtlety due to the artificial
fourth root of the fermion determinant necessary for the staggered
fermions, for which the Feynman-Hellmann theorem is modified.
Their result appears to be slightly higher than ours.
\section{Conclusions}
\label{Sec:Conclusion}

In this paper we calculate the nucleon strange quark content on the
lattice directly from the nucleon three-point function in two-flavor QCD. 
Chiral symmetry is exactly preserved by employing the overlap fermion
formulation on the lattice.
This is crucial in the calculation of the strange quark content in
order to avoid large contaminations from the operator mixing effects,
that were missing in many of the previous calculations.

The lattice calculation of the disconnected diagram is technically
challenging. 
In this work we attempted various options of the all-to-all propagator
technique and the low-mode averaging together with the source and sink
smearings.
By optimizing those, we could finally obtain the nonzero signal at each
quark mass; the value extrapolated to the physical quark masses is
away from zero by 1.5 standard deviation.

The results for $f_{T_s}$ and $y$ are in good agreement with our
previous estimate using the spectrum method \cite{Ohki:2008ff}, 
and favor small strange quark content $y\approx 0.05$,
which is an order of magnitude smaller than previous lattice
calculations without respecting chiral symmetry, 
which we now believe unreliable. 

For more realistic calculations, we must include the dynamical strange
quark in the simulation.
Such a calculation is already underway using both the spectrum and
direct methods \cite{Takeda:2009ga,Ohki:2009mt}.
It is also interesting to extend this study to other baryon observables
involving disconnected quark loops, such as the strange quark spin
fraction of the nucleon. 

\begin{acknowledgments}
  Numerical simulations are performed on Hitachi SR11000 and 
  IBM System Blue Gene Solution 
  at the High Energy Accelerator Research Organization (KEK) 
  under the support of its Large Scale Simulation Program (No.~09-05).
  This work is supported in part by the Grant-in-Aid of the
  Ministry of Education (No.~20105001, No.~20105002, No.~20105003,
  No.~20105005, No.~20340047, and No.~21684013). 
\end{acknowledgments}

\end{document}